\documentclass[aps,
twocolumn,
raggedbottom,
prl,
showpacs,
nobalancelastpage,
groupedaddress]{revtex4}
\usepackage{graphicx}
\usepackage{amsmath}
\usepackage{amsfonts}
\usepackage{amssymb}
\usepackage{hyperref}
\usepackage{units}

\newcommand{\da}{\dagger}
\newcommand{\ve}[1]{\ensuremath{\mathbf{#1}}}
\newcommand{\aver}[1]{\langle #1 \rangle}
\newcommand{\cre}[2]{\ensuremath{#1_{#2}^\da}}
\newcommand{\ann}[2]{\ensuremath{#1_{#2}^{ }}}
\newcommand{\abso}[1]{\lvert #1 \rvert}
\newcommand{\p}{\prime}
\newcommand{\com}[1]{[#1]}

\newcommand{\btau}{\boldsymbol{\tau}}
\newcommand{\brho}{\boldsymbol{\rho}}

\newcommand{\vG}{\ve{G}}

\allowdisplaybreaks[1]

\begin{document}

\title{Helicity and electron correlation effects on transport
  properties of double-walled carbon nanotubes}

\author{Shidong Wang and Milena Grifoni} \affiliation{Theoretische
  Physik, Universit\"at Regensburg, 93040 Regensburg, Germany.}

\date{\today}

\begin{abstract}
  We analytically demonstrate helicity determined selection rules for
  intershell tunneling in double-walled nanotubes with commensurate
  (c-DWNTs) and incommensurate (i-DWNTs) shells.  For i-DWNTs the
  coupling is negligible between lowest energy subbands, but it
  becomes important as the higher subbands become populated. In turn
  the elastic mean free path of i-DWNTs is reduced for increasing
  energy, with additional suppression at subband onsets.  At low
  energies, a Luttinger liquid theory for DWNTs with metallic shells
  is derived.  Interaction effects are more pronounced in i-DWNTs.
\end{abstract}


\maketitle

Due to their unusual physical properties, cf. e.g.~\cite{saito:1998}
carbon nanotubes have attracted lots of attention.  Carbon nanotubes
can be single-walled (SWNT) or multi-walled (MWNT), depending on
whether they consist of a single or of several graphene sheets wrapped
onto coaxial cylinders, respectively. Electronic properties of SWNTs
are mostly understood~\cite{saito:1998}. In particular, SWNTs are
usually ballistic conductors \cite{white:nature1998}, and whether a
SWNT is metallic or semiconducting is solely determined by its so
called chiral indices $(n,m)$.  Due to the one-dimensional character
of the electronic bands at low energies, Luttinger liquid features at
low energies have been predicted \cite{egger:prl1997, kane:prl1997}
and observed \cite{yao:nature1999, bockrath:nature1999}.  The
situation however, is much less clear for MWNTs. Except for few
experiments, see e.g.~\cite{frank:science1998,urbina:prl2003}, MWNTs
are typically diffusive conductors, see e.g.~\cite{langer:prl1996,
  bachtold:nature1999}, with current being carried by the outermost
shell at low bias ~\cite{bachtold:nature1999, fujiwara:prb1999} and
also by inner shells at high bias~\cite{collins:prl2001}. Intershell
conductance measurements consistent with tunneling through orbitals of
nearby shells have recently been reported \cite{bourlon:prl2004}.
Moreover, which kind of electron-electron correlation effects
determine the observed zero-bias
anomalies~\cite{bachtold:prl2001,graugnard:prb2001} of MWNTs is still
under debate.  To better understand these features, i.e., the role of
{\em inter}shell coupling on transport properties of MWNTs, some
experimental~\cite{kociak:prl2002} and theoretical
\cite{saito:jap1993,kwon:prb1998,lambin:prb2000, triozon:prb2004,
  ahn:prl2003,uryu:prb2004,chen:jpcm2005} works focussed on the
simplest MWNT's realization, namely on double-walled nanotubes
(DWNTs).  One main outcome is that a relation must exist between the
intershell coupling, shell helicity and transport properties.
Specifically, two shell are called commensurate (incommensurate), if
the ratio between their respective unit cell lengths along the tube
axis, is rational (irrational)~\cite{saito:1998}.  For example, using
tight-binding models, Saito {\em et al.}~\cite{saito:jap1993}
numerically found energy gaps opened by the intershell coupling in a
DWNT with two armchair (and hence commensurate) shells. Ab-initio
calculations~\cite{kwon:prb1998, lambin:prb2000} confirmed these
results. In general, numerical evidence of a negligible intershell
coupling in DWNTs with incommensurate shells (i-DWNT) at low energies
is found ~\cite{saito:jap1993, lambin:prb2000, triozon:prb2004,
  ahn:prl2003}.

In this Letter, we derive an analytical expression, yielding {\em
  helicity-dependent selection rules} for tunneling, for the effective
intershell coupling. For i-DWNTs the intershell coupling is negligible
between the {\em lowest} subbands but it becomes important when {\em
  higher} subbands are involved. We show that this in turn yields an
elastic mean free-path which decreases with energy and which shows a
characteristic suppression at subbands onset. Then by including intra-
and inter-shell Coulomb interactions, we show that metallic DWNTs can
be described by Luttinger liquid theory at low energies. The tunneling
density of states has a power-law behavior with different exponents
for i-DWNTs and commensurate-shells DWNTs (c-DWNTs).

\begin{figure}[htbp]
  \includegraphics[width=4cm]{dwnt-fig1a.eps}
  \includegraphics[width=3.8cm]{dwnt-fig1b.eps}
  \caption{
    (a) Graphene lattice. The $x$ and $y$ axes are along the armchair
    and zigzag axes respectively.  The distance between two nearest
    carbon atoms is $a_0 \sim \unit[1.42]{\AA}$. The unit lattice
    vectors are $\ve{a}_1$ and $\ve{a}_2$. The vectors $\ve{d}_i$
    connect three nearest neighbour atoms, while $\brho$ and $\btau$
    are two vectors required to specify the position of a carbon atom.
    (b) Cross section of a DWNT. Atoms $A$ and $B$ in two shells of
    radii $R_a$ and $R_b$, respectively, are projected onto this cross
    section.  }
  \label{fig:graphene}
\end{figure}

To derive the helicity-dependent selection rules, we use a
tight-binding model for non interacting electrons with one
$\pi$-orbital per carbon atom ~\cite{saito:1998} and follow
~\cite{maarouf:prb2000}.  This model is described by the Hamiltonian
\begin{equation}
  H_0 = \sum_{\beta} \sum_{\aver{ij}} \gamma_0 \cre{c}{\beta i}
  \ann{c}{\beta j} + \sum_{ij} t_{\ve{r}_{ai}, \ve{r}_{bj}}
  \cre{c}{ai} \ann{c}{bj} + \mathrm{H.c.},
\end{equation}
where $\beta = a,b$ is the shell index, $\aver{ij}$ is a sum over
nearest neighbors in a shell, $\gamma_0 \sim
\unit[2.7]{eV}$~\cite{saito:1998} is the intrashell nearest neighbour
coupling.  The intershell coupling is $t_{\ve{r}_{ai}, \ve{r}_{bj}} =
t_0 e^{ -(d(\ve{r}_{ai}, \ve{r}_{bj}) - \Delta) /a_t}$, where $t_0\sim
\unit[1.1]{eV}$, $\Delta \sim \unit[0.34]{nm}$, $d(\ve{r}_{ai},
\ve{r}_{bj})$ is the distance between two atoms, and $a_t \sim
\unit[0.5]{\AA}$~\cite{saito:1998}.  We introduce the transformation
\begin{equation}
  \ann{c}{\beta j} = \frac{1}{\sqrt{N_\beta}}\sum_{\ve{k}} e^{i \ve{k} \cdot
    \ve{r}_j} \ann{c}{\beta \eta (j) \ve{k}}\;,
\end{equation}
where $\eta =\pm $ is the index for the two interpenetrating
sublattices in a graphene sheet, and $N_{\beta}$ is the number of
graphene unit cells on a shell. The Hamiltonian reads
\begin{eqnarray}
  H_0  &=&  \sum_{\beta \eta \ve{k}} \gamma_{\ve{k}}
 c_{\beta \eta \ve{k}}^\da c_{\beta -\eta \ve{k}}\nonumber \\
   &+&
  \sum_{ \ve{k}_a\ve{k}_{b}}\sum_{\eta_a\eta_{b}}
 \mathcal{T}_{\eta_a\eta_{b}}(\ve{k}_a,\ve{k}_{b})\cre{c}{a\eta_a\ve{k}_a}
  \ann{c}{b\eta_{b}\ve{k}_{b}} + \mathrm{H.c.}  \;,
\label{eq:H-plane}
\end{eqnarray}
where the {\em intra}shell coupling is $\gamma_\ve{k} = \sum_{j=1}^3
\gamma_0 e^{i\ve{k}\cdot\ve{d_j}}\equiv |\gamma_\ve{k}|s$, with
$\ve{d}_j$ the vectors connecting the three nearest neighbour carbon
atoms. Introducing ${\bf r}_{\beta}={\bf R}+{\bf X}_{\beta}$, with
${\bf R}$ a lattice vector, ${\bf X}_{\beta}=\brho_\beta+\eta_\beta
\btau$, where $\brho_a - \brho_b$ describes the relative position of
the two shells, cf. Fig. 1, the elements of the {\em inter}shell
$2\times 2$ coupling matrix are
\begin{equation}
  \label{eq:effective-coupling}
\mathcal{T}_{\eta_a\eta_b}(\ve{k}_a, \ve{k}_b) =
  \sum_{\ve{G}_a\ve{G}_b}
    e^{i\ve{G}_a\cdot \ve{X}_a  -
    i\ve{G}_b\cdot \ve{X}_b} t_{\ve{k}_a+\ve{G}_a,
    \ve{k}_b+\ve{G}_b}\;.
\end{equation}
Here $\ve{G}$ is the graphene reciprocal lattice vector
\begin{equation*}
  \vG 
= \frac{4\pi}{3a_0}
  \Bigl(\frac{\sqrt{3}}{2} (l_1 - l_2),\; \frac{1}{2}(l_1 + l_2)
  \Bigr)\;,
\end{equation*}
with $l_1, l_2 = 0, \pm 1, \pm 2, \dots$.  Finally,
\begin{equation*}
 t_{\ve{q}_a, \ve{q}_b} =
    \frac{1}{A_{\mathrm{cell}}^2\sqrt{N_aN_b}} \int d\ve{r}_a d\ve{r}_b
    e^{i (\ve{q}_b\cdot\ve{r}_b - \ve{q}_a\cdot\ve{r}_a)}
    t_{\ve{r}_a, \ve{r}_b},
\end{equation*}
with $A_{\mathrm{cell}}$ the area of a graphene unit cell.
We notice that the Hamiltonian Eq.~\eqref{eq:H-plane} is diagonalized,
in the absence of intershell coupling, by the transformation
$U=\tfrac{1}{\sqrt 2}
\bigl(\begin{smallmatrix} s & s \\
  -s^*& s^*\end{smallmatrix} \bigr)$. For later purposes, cf.
Eq.~\eqref{eq:T-matrix}, we call $\tilde{\mathcal{T}}_{\nu_a\nu_b} =
(U^\dagger {\cal T}U)_{\nu_a\nu_b}$ the elements of the intershell
tunneling matrix between two Bloch states in different shells. Here
$\nu=\mp $ is the index for bonding/ antibonding states corresponding
to negative/positive energies $\varepsilon_{\beta,\nu}(\ve{k})$ with
$\beta=a,b$, respectively.

\begin{figure}[htbp]
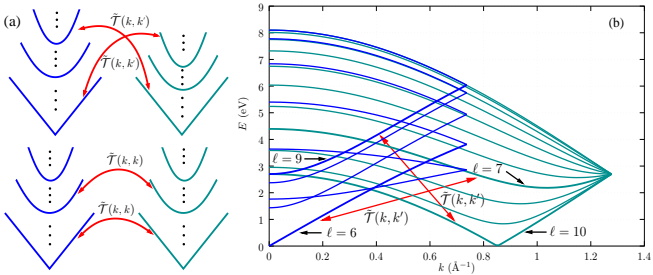

  \begin{minipage}[c]{3.0cm}
    \includegraphics[width=3.0cm]{dwnt-fig2Ua.eps} \\\vspace{0.1cm}
    \includegraphics[width=3.0cm]{dwnt-fig2Da.eps}
  \end{minipage}
  \begin{minipage}[c]{5.5cm}
    \includegraphics[width=5.5cm,height=3.6cm]{dwnt-fig2b.eps}
  \end{minipage}
  \caption{
    (a) Schematics of the effect of the selection rules for i-DWNT
    (upper) and c-DWNT (down).  (b) Energy subbands of the i-DWNT
    $(9,0)@(10,10)$ in the absence of intershell coupling. The effect
    of the latter is to induce transitions between different subbands
    in different shells. The dominant coupling of the lowest armchair
    (zig-zag) subband is indicated. }
  \label{fig:Band-structure}
\end{figure}

It is convenient to introduce coordinates $u$ and $v$, which are along
the tube axis and the circumference direction respectively, cf.
Fig.~\ref{fig:graphene}(b). Then $\ve{k}_a\cdot\hat{\ve{v}} = k_{va}$
obeys $\ell_{a} = k_{va}R_a$, due to the periodic boundary conditions
along the circumference. Likewise, $\ell_b = k_{vb}R_b$. Here, the
integers $\ell_{a}$ and $\ell_b$ characterize energy subbands. In
contrast, $k_u = \ve{k}\cdot\hat{\ve{u}}$ is continuous, cf.
Fig.~\ref{fig:Band-structure}(b).  The distance between two atoms $A =
(R_a \cos (v_a/R_a), R_a \sin (v_a/R_a), u_a)$ and $B = (R_b \cos
(v_b/R_b), R_b \sin (v_b/R_b), u_b)$ is then $d(\ve{r}_a , \ve{r}_b)
\equiv D(v_a/R_a-v_b/R_b, u_a-u_b)$ where
\begin{equation*}
  D(z_1, z_2) =
  \sqrt{ \abso{R_a - R_b}^2 + 4R_aR_b
   \sin^2\left(z_1/2 \right) +
 (z_2)^2}.
\end{equation*}
We then find $t_{\ve{q}_a, \ve{q}_b} = t \delta(q_{va}R_a - q_{vb}R_b)
\delta(q_{ua} - q_{ub})$, where the prefactor $t$ is calculated as
\begin{equation*}
  \begin{split}
    t =t_0 \int &\frac{dz_1 dz_2 \;e^{-(D(z_1,
        z_2) - \Delta)/a_t}}{A_{\mathrm{cell}}^2\sqrt{N_aN_b}} \\
    & \times e^{i z_1(q_{vb}R_b + q_{va}R_a)} e^{i z_2(q_{ub} +
      q_{ua})}.
  \end{split}
\end{equation*}
Therefore, according to the two $\delta$-functions in $t_{\ve{q}_a,
  \ve{q}_b}$, the effective intershell couplings
$\mathcal{T}_{\eta_a\eta_b}(\ve{k}_a, \ve{k}_b )$ between two shells
$(n_a, m_a)$ and $(n_b, m_b)$ are nonzero if they satisfy the
following {\em selection rules},
\begin{subequations}
  \label{eq:selection-rule}
\begin{align}
\label{eq:v-d}
\ell_a + (n_a l_{1a} + m_{a} l_{2a}) &=
\ell_b + (n_b l_{1b} + m_{b} l_{2b}), \\
k_{ua} + \mathcal{F}(n_a, m_a) &= k_{ub} + \mathcal{F}(n_b, m_b),
\label{eq:u-d}
\end{align}
\end{subequations}
with $\mathcal{F}(n,m)= \tfrac{2\pi}{3a_0\mathcal{L}(n, m)} \bigl((n +
2m)l_{1} - (2n + m) l_{2}\bigr).$ Here $\sqrt{3}a_0\mathcal{L}(n,m)$,
with $\mathcal{L}(n,m)= \sqrt{n^2+m^2+nm}$, is the circumferential
length of shell $(n,m)$.  At low energies only the lowest subband
determined by $3\ell_{\beta} = 2n_\beta + m_\beta$ in each shell is
important, which {\em fixes} the values $\ell_a$ and $\ell_b$.  For
c-DWNTs, e.g. if the two shells are either both armchair or zig-zag,
Eqs.~\eqref{eq:selection-rule} can always be satisfied. Moreover, the
dominant contribution is for $k_{ua}=k_{ub}$.  On the other hand, for
i-DWNTs, e.g. a $(9,0)@(10,10)$, the selection rule Eq.~\eqref{eq:u-d}
can only be satisfied if the difference $ k_{ua}-k_{ub}$ takes finite
values.  At low energies, this condition is never met. At higher
energies higher subbands must be considered as well, and the selection
rules can be satisfied.  Notice that whenever the l.h.s. and r.h.s. of
Eqs.~\eqref{eq:selection-rule} are not close to zero, the effective
intershell coupling is exponentially suppressed~\cite{note:1}.  We
show now that, for i-DWNTs, the increase of the inter-shell tunneling
is at the origin of an elastic mean-free path $l_{el} $ which
decreases with increasing energy, and which shows a characteristic
suppression at each subband onset. Our analytical results are in
agreement with recent ab-initio calculations, showing a cross-over
from ballistic to diffusive behavior in DWNT as the energy increases
\cite{triozon:prb2004, ahn:prl2003}, as well as with the experimental
observation that MWNT mostly exhibit diffusive behavior. To evaluate
the elastic mean-free path $l_{el,b}(E)=v_F \tau_b (E)$ for electrons
in the shell $b$, the life-time $\tau_b (E)$ for electrons with energy
$E$ is needed.  Here $v_F~ = \unit[8\times 10^5]{m/s}$ is the Fermi
velocity for nanotubes. To be definite, $\hbar /\tau_b
(E)=\sum_{\ve{k}, \nu=\pm} (\hbar /\tau_{b,\ve{k}\nu})\delta
(E-\varepsilon_{b,\nu}(\ve{k}))$, with $\varepsilon_{b,\nu}(\ve{k})$
the dispersion relation in shell $b$, and
\begin{equation}
  \frac{\hbar}{\tau_{b,\ve{k}\nu}}  = \sum_{\beta=a,b \atop \nu^\p=\pm}  \int
  dk_u^\p \, 
  \abso{T_{\ve{k}\ve{k}^\p,\nu\nu^\p}^{b\beta}
    \bigl(\varepsilon_{b,\nu}(\ve{k}) \bigr)}^2
    \delta(\varepsilon_{b,\nu}(\ve{k}) -
    \varepsilon_{\beta,\nu^\prime}(\ve{k}^\p)). 
\label{eq:tau-k}
\end{equation}
For DWNTs the $4 \times 4$ $T$-matrix \cite{mahan:1990} is evaluated
to be
\begin{equation*}
  T_{\ve{k}\ve{k}^\p}(\omega) = \mathcal{V}(\ve{k},\ve{k}^\p) + \sum_{\ve{k}_1}
  \mathcal{V}(\ve{k},\ve{k}_1)G(\ve{k}_1,\omega)T_{\ve{k}_1\ve{k}^\p}(\omega),
\end{equation*}
where $\mathcal{V}^{ba}_{\nu_a\nu_b}(\ve{k},\ve{k^\prime}) =
\bigl(\mathcal{V}^{ab}_{\nu_a\nu_b}(\ve{k},\ve{k^\prime})\bigr)^\star
= \tilde{\mathcal{T}}_{\nu_a\nu_b}(\ve{k},\ve{k}^\p)$ and
$\mathcal{V}^{\beta\beta}_{\nu\nu^\prime}(\ve{k},\ve{k^\prime}) = 0$.
The elements of the retarded Green's function $G$ are
$G^{\beta\beta^\prime}_{\nu\nu^\prime}(\omega,\ve{k}) = (\omega -
\varepsilon_{\beta,\nu}(\ve{k}) + i0^+)^{-1}
\delta_{\beta\beta^\prime} \delta_{\nu\nu^\prime}.$ For i-DWNTs
$\tilde{\mathcal{T}}_{\nu_a\nu_b}$ couples subbands with different
energies, cf. Fig. 2. Thus, in general is
$\varepsilon_{b,\nu}(\ve{k})\neq \varepsilon_{a,\nu^\prime}
(\ve{k}^\prime)$, i.e., $\delta (\varepsilon_{b,\nu}(\ve{k})-
\varepsilon_{a,\nu^\prime} (\ve{k}^\prime))=0$, and $\beta=b$ in
Eq.~\eqref{eq:tau-k}. Hence, to lowest order in ${\cal \tilde T}$, and
if $\varepsilon_{b,\nu}(\ve{k})\neq \varepsilon_{a,\nu^\prime}
(\ve{k}^\prime)$, the life-time $\tau_{b,\ve{k}\nu}$ is obtained
inserting in Eq.~\eqref{eq:tau-k}
\begin{equation}
  \label{eq:T-matrix}
  T^{bb}_{\ve{k}\ve{k}^\p,\nu\nu^\prime}(\varepsilon_{b,\nu}(\ve{k})) =
  \sum_{\ve{k}_1 ,\nu_1=\pm} 
  \frac{\tilde{\mathcal{T}}_{\nu\nu_1}^*(\ve{k},
    \ve{k}_1)\tilde{\mathcal{T}}_{\nu_1\nu^\prime}(\ve{k}_1,
      \ve{k}^\p)}{\varepsilon_{b,\nu}(\ve{k}) -  
    \varepsilon_{a,\nu_1}(\ve{k}_1) + i0^+}\;.
\end{equation}
The elastic mean free paths $l_{el,(10,10)}$ and $l_{el,(9,0)}$ for
electrons in the outer and inner shell, respectively, of a
(9,0)@(10,10) DWNT are shown in Fig. 3. It is clearly shown that {\em
  before} the first subband onset, the motion is ballistic also for
i-DWNTs of lengths up to $\simeq 5\mu m$.

In the remaining of the paper we consider only metallic shells and
include electron-electron correlation effects in the low energy regime
where only the first subband of each shell is populated. In this
regime transport is ballistic for c-DWNTs as well as for i-DWNTs. Due
to the linearity of the dispersion relation, a multi-channel Luttinger
liquid description can be used~\cite{matveev:prl1993,egger:prl1999}.
\begin{figure}[htbp]
  \includegraphics[width=8.0cm]{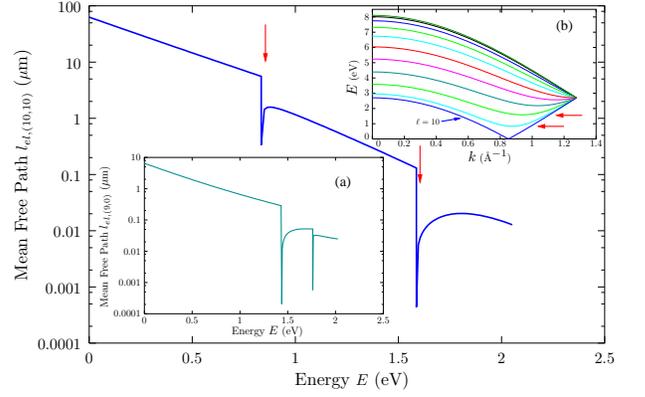}
  \caption{Elastic mean free paths
    for electrons in the outer and inner (inset (a)) shells of the
    i-DWNT $(9,0)@(10,10)$.  Notice the dips in correspondence of the
    first two subband onsets.}
 \label{fig:mean-free-path}
\end{figure}

At first, we consider an i-DWNT where the intershell coupling can be
ignored. The unperturbed Hamiltonian can be written as
\begin{equation}
  \label{eq:H-incommensurate}
  H_0 = -i \hbar v_F \sum_{r\alpha\sigma\beta} r \int du\,
  \cre{\psi}{r\alpha\sigma\beta} \partial_u
  \ann{\psi}{r\alpha\sigma\beta}\;,
\end{equation}
where $r=\pm$ is the index for right/left movers, $\alpha = \pm$ for
the two independent Fermi points of a shell, and $\sigma = \pm$ for up
and down spins. The electron density operator is $ \rho_{\beta} (u) =
\sum_{r\alpha\sigma\beta} \cre{\psi}{r\alpha\sigma\beta}(u)
\ann{\psi}{r\alpha\sigma\beta}(u)$.  The two shells are only coupled
by the Coulomb interaction, which gives rise to forward, backward, and
Umklapp scattering processes.  Experimentally, the Fermi points of
nanotubes are usually shifted away from the half-filling due to doping
or external gates. We assume that this is the case and hence neglect
Umklapp processes.  Since we are not interested in the extremely low
temperature case, the backward scattering processes are also ignored
here~\cite{egger:prl1999}. In the following, we only consider forward
scattering processes described by the Hamiltonian
\begin{equation}
  \label{eq:U}
  H_{\mathrm{FS}} = \frac{1}{2} \sum_{\beta\beta^\p}\int du du^\p \;
  \rho_{\beta}(u) V_{\beta\beta^\p}(u - u^\p) \rho_{\beta^\p}(u^\p),
\end{equation}
with the effective one dimensional interaction
\begin{equation*}
  V_{\beta\beta^\p}(u-u^\p) = \int_0^{2\pi R_{\beta}} \int_0^{2\pi
    R_{\beta^\p}} \frac{dv dv^\p}{(2\pi)^2R_{\beta}R_{\beta^\p}} \;
  U_{\beta\beta^\p}(\ve{r} - \ve{r}^\p)\;,
\end{equation*}
where $U({\ve{r}})$ is the Coulomb interaction. The Hamiltonian $H=H_0
+ H_{\mathrm{FS}}$ can be diagonalized by the bosonization procedure
discussed in Ref.~\cite{matveev:prl1993}.  We introduce bosonic field
operators for the total/relative $(\delta =\pm)$ charge/spin $(j=c,s)$
modes in shell $\beta$, as well the total/relative $(\xi = \pm)$ modes
with respect to the two shells obeying the commutation relation
$\com{\Theta_{j\delta\xi}(u), \phi_{j^\p\delta^\p\xi^\p}(u^\p)} =
-(i/2) \delta_{jj^\p} \delta_{\delta\delta^\p} \delta_{\xi\xi^\p}
\mathrm{sgn}(u-u^\p)$.  The Hamiltonian $H$ can be then decoupled into
8 modes as
\begin{equation}
  \sum_{j\delta\xi} \frac{\hbar v_{j\delta\xi}}{2} \int du \biggl(
  K_{j\delta\xi} \bigl(\partial_u \Theta_{j\delta\xi}(u) \bigr)^2 +
  \frac{1}{K_{j\delta\xi}} \bigl(\partial_u \phi_{j\delta\xi}(u) \bigr)^2
  \biggr).
\end{equation}
Only the two total charge modes are renormalized by the Coulomb
interactions with velocities $v_{c+\pm}=v_F/K_{c+\pm}$ and interaction
parameters
\begin{equation}
  \label{eq:parameters}
 \frac{1}{K^2_{c+\pm}} = 1 +\frac{2}{\hbar\pi v_F}
 \Bigl((\tilde{V}_{aa} + \tilde{V}_{bb})\pm
    \sqrt{ (\tilde{V}_{aa} - \tilde{V}_{bb})^2
    + \tilde{V}_{ab}^2\;\;}\Bigr) \;,
\end{equation}
where $\tilde V_{\beta\beta^\p}=\tilde V_{\beta\beta^\p }(2\pi/L)$ is
the Fourier transform of the interaction potential at long-wave
lengths, with $L$ the nanotube length. The remaining modes are neutral
with parameters $v_{j\delta\xi} = v_F$ and $K_{j\delta\xi} = 1$.

We consider now c-DWNTs where intershell tunneling is relevant.  The
intershell tunneling Hamiltonian is
\begin{equation}
H_t =   \sum_{r\alpha\sigma}  \tilde{\mathcal{T}}_{0} \int du\,
 \cre{\psi}{r\alpha\sigma a}(u) \ann{\psi}{r\alpha\sigma b}(u) +
 \mathrm{H.c} \;,
\end{equation}
where for simplicity the tunneling element
$\tilde{\mathcal{T}}_{++}(\ve{k},\ve{k}^\prime)$ is evaluated at
$\ve{k}=\ve{k}^\prime=\ve{K}$ with the Fermi point $\ve{K}$ of
graphene, and is the constant ${\cal \tilde T}_{0}$. As detailed in
\cite{egger:prl1999}, the Hamiltonian $H_0+H_t$ can be exactly
diagonalized. One finds the same form as in
Eq.~\eqref{eq:H-incommensurate} where now the index $\beta = 0, \pi$
stands for bonding and anti-bonding states, respectively.  Moreover,
the Fermi wave vectors of the two independent Fermi points are shifted
as $ k^{\alpha}_{F} \longrightarrow k^{\alpha}_{F} \pm ({\cal \tilde
  T}_0/\hbar v_F)$ where $\pm$ stand for $\pi$ and $0$, respectively.
We retain again only (intraband and interband) forward scattering
described by the Hamiltonian Eq.~\eqref{eq:U}, where now the
scattering potentials $\tilde V_{\beta\beta^\prime}$ are
$\tilde{V}_{00} = \tilde{V}_{\pi\pi} = \tilde{V}_{0\pi}/2 =
(\tilde{V}_{aa} + \tilde{V}_{bb} + \tilde{V}_{ab})/4$, so that
bosonization brings again the total Hamiltonian in the form
Eq.~\eqref{eq:parameters} with 6 neutral modes and 2 renormalized
total charge modes.  The tunneling density of states (TDOS) of both
shells, $\rho_{b/a}(\varepsilon)$, immediately follows \cite{note:2}.
For i-DWNT is
\begin{equation*}
\rho_{b/a}(\varepsilon) \sim
\abso{\varepsilon}^{\alpha_{b/a}}\;,
\end{equation*}
with exponents $\alpha_{b/a}$ being different for electrons tunnelling
into the middle or end of a nanotube:
\begin{equation}
\label{eq:aend-out}
  \begin{split}
    \alpha_{\mathrm{end}, b/a} &= \frac{1}{4} \sum_{\xi =\pm} A_{b/a}
    \biggl(\frac{1}{K_{c+\xi}} - 1 \biggr) , \\
    \alpha_{\mathrm{bulk}, b/a} &= \frac{1}{8} \sum_{\xi=\pm} A_{b/a}
    \biggl( K_{c+\xi} + \frac{1}{K_{c+\xi}} - 2\biggr)\;.
  \end{split}
\end{equation}
Here the coefficients $A_{a}$, $A_{b} = 1- A_{a}$ are related to the
eigenvalue problem \cite{egger:prl1999, matveev:prl1993}.  For c-DWNT
is
\begin{equation*}
\rho_{b/a}(\varepsilon) \sim
\abso{\varepsilon}^{\alpha_{0}} +
\abso{\varepsilon}^{\alpha_\pi}\Theta(\epsilon - 2{\cal \tilde
T}_0),
\end{equation*}
where $\Theta(x)$ is the Heaviside step function and $2{\cal\tilde
  T}_0$ is the gap between antibonding and bonding states. Because the
intraband forward scattering potentials are equal, is $\alpha_0 =
\alpha_{\pi}$. We find $\alpha_{\mathrm{end/bulk}}$ given by
Eq.~\eqref{eq:aend-out} with $A_{b} = A_{a}=1/2$.  For illustration we
calculate the tunneling exponents for the $(10,10)$ shell of a
$(9,0)@(10,10)$ and of a $(5,5)@(10,10)$ with radii $R_{a} \approx
\unit[3.4]{\AA}$ and $R_{b} \approx \unit[6.8]{\AA}$. We find
$\alpha_{\mathrm{end}} = 1.21, \alpha_{\mathrm{bulk}} = 0.50$ for a
$(9,0)@(10,10)$ DWNT and $ \alpha_{\mathrm{end}} = 0.80$,
$\alpha_{\mathrm{bulk}} = 0.34$ for a $(5,5)@(10,10)$ DWNT. For
comparison, for a $(10,10)$ SWNT is $\alpha_{\mathrm{end}} = 1.25$ and
$\alpha_{\mathrm{bulk}} = 0.52$. Hence, the exponents of DWNTs
decrease due to the screening effect of the inner shell with respect
to a SWNT. The intershell coupling reduces the exponents further.
Notice that for Fermi liquids is $\alpha_{\mathrm{end/bulk}} = 0$.

In summary, we derived selection rules according to which the
intershell coupling is only negligible in i-DWNTs at low energies. An
analytical expression in Born-approximation for the elastic mean free
path was provided.  Including the Coulomb interaction, we developed a
low energy Luttinger liquid theory for metallic DWNTs according to
which the intershell coupling strongly reduces the tunneling density
of state exponents in c-DWNT with respect to those of i-DWNTs.

\begin{acknowledgments}
  The authors would like to thank G. Cuniberti, J. Keller and C.
  Strunk for useful discussions.
\end{acknowledgments}


\begin{thebibliography}{42}
\bibitem{note:1} This holds, e.g. for finite length DWNTs, where
  conservation law for the wave vectors along the tube axis is broken
  but the selection law for the wave vectors along the circumference
  still holds.
\end{thebibliography}

\end{document}